\documentclass[12pt]{article}
\def\be{\begin{equation}}
\def\ee{\end{equation}}
\def\ba{\begin{eqnarray}}
\def\ea{\end{eqnarray}}
\def\la{\langle}
\def\ra{\rangle}

\def\h{\hskip 1cm}

\usepackage{graphicx}
\begin{document}
\begin{titlepage}
\vspace{4cm}

\begin{center}{\Large \bf A complete characterization of the spectrum of the Kitaev model on spin ladders}\\
\vspace{2cm}\h Vahid Karimipour \footnote{vahid@sharif.edu}\\
\vspace{1cm} Department of Physics, Sharif University of
Technology,\\
 11155-9161, Tehran, Iran\\

\end{center}
\begin{abstract}
We study the Kitaev model on a ladder network and find the complete
spectrum of the Hamiltonian in closed form.  Closed and manageable
forms for all eigenvalues and eigenvectors, allow us to calculate
the partition function and averages of non-local operators in
addition to the reduced density matrices of different subsystems at
arbitrary temperatures. It is also briefly discussed how these
considerations can be generalized to more general lattices,
including three-leg ladders and two dimensional square lattices.
\end{abstract}
 \hspace{.3in}
\end{titlepage}

\section{Introduction}\label{Intro}

There are very few exactly solvable spin systems, in the sense that
their complete spectra can be determined in closed form. The most
notable one is the Ising model in transverse field or the XY model
\cite{Mattis}, where the spectrum can only be determined with the
help of highly nonlocal Jordan-Wigner transformation, rendering the
calculation of correlation functions very difficult. For the other
example, the Heisenberg spin chain, the energy eigenvalues and
eigenvectors can only be determined implicitly, by solving the
system of coupled non-linear Bethe ansatz equations \cite{Bethe,
Baxter}. For a large class of spin models, only the ground state can
be found by the matrix product approach
\cite{FCS1,Zittartz1,Zittartz2,AKS1,AKM,KM,ABK}. Having an exactly
solved quantum many body system, in the sense of complete
determination of its spectrum, is always a fortunate situation,
which enables one to make a detailed study of thermal and dynamical
properties. It also allows one to solve other related systems by
perturbation techniques. \\

While the study of many body systems has been traditionally done in
the community of condensed matter and statistical physics, and also
by mathematical physicists interested in exact solutions, in recent
years, these systems have attracted a lot of attention from the
quantum information community. The reason is at least two-fold: on
the one hand an array or lattice of two-level quantum systems
(qubits) is the natural candidate for implementation of quantum
information processing tasks and on the other hand, concepts and
tools developed in quantum information \cite{Nielsen}, which are
mainly aimed at characterizing the nature of quantum states like
their entanglement, have been quite useful in understanding
different phenomena in such systems, i.e. quantum phase transitions
\cite{osterloh, osborne}. \\

While in condensed matter physics, the focus is on the Hamiltonian
and interactions, in quantum information, the emphasis is on the
quantum states and their properties, i.e. their bi-partite and
multi-partite entanglement. Needless to say, we are not always in
the happy situation to have both a physically plausible Hamiltonian
on the one hand and an easily obtainable spectrum on the other. For
example in the matrix product formalism, although we can construct
ground states with desired symmetries, it is not guaranteed that the
parent Hamiltonians are of real experimental interest. Fortunately
with the recent advances in optical lattices and cold atoms, we have
more freedom in manipulating systems of many body two-level systems
(qubits). Therefore there is less reservation than before in
proposing many-body Hamiltonians whose ground states, or the low
level excited states, may have desirable properties for
implementation of quantum information processing. \\

In this regard, an interesting model has been recently proposed
\cite{KitaevBasic1} which has the very desirable feature of showing
topological order and anyonic excitations. This was the first model
for topological implementation of quantum computation, where qubits
are encoded into the homological classes of loops on a surface which
are hence resistent to local errors. The important point is that
topological order is not related to the symmetries of the
Hamiltonian and it is robust against arbitrary local perturbations,
even those that destroy all the symmetries of the Hamiltonian. \\

On any lattice, the Kitaev model is defined by the Hamiltonian
\begin{equation}\label{Hamiltonian}
    H = -J\sum_sA_s-K\sum_p B_p,
\end{equation}
where $J$ and $K$ are positive coupling constants, and the vertex
operators $A_s$, and plaquette operators $B_p$ are constructed from
Pauli operators $X=\sigma_{x}$ and $Z=\sigma_z$ as follows: $A_s$ is
the product of all $X$'s on the links shared by the vertex $s$ and
$B_p$ is the product of all $Z$ operators around a plaquette. These
operators commute with each other for any
geometry of the network.\\

It is well-known that the ground state of this model has symmetries
not inherited from the Hamiltonian but from the topology of the
surface on which the model is defined. On a genus $g$ surface
without boundary, the ground state of model (\ref{Hamiltonian}) has
$4^g$ fold degeneracy with a gap which cannot be removed by local
perturbations. Therefore on such a surface, the ground space can
encode $2g$ qubits, in a way where resistance to errors is
automatically ensured by the topology. In recent years, there has
been intensive activity on this model and its variations and
generalizations in many directions, see the works
 \cite{Preskill, Delgado1,Delgado2, Pachos1,Freedman1,Freedman2,Honeycomb1,Honeycomb2,Kitaev1D,Castelnovo}
and references therein for a sample.  The ground states of the
Kitaev model can be characterized rather simply in a formal way.
Since $A_s^2=B_p^2=I$, the ground state is one which is stabilized
by all these operators, i.e. $A_s|g.s\ra=B_p|g.s\ra=|g.s\ra$. One
such state can be written as $|g.s\ra=\sum_{C} C_z|\Omega\ra$, where
$|\Omega\ra$ is a sea of spins in $+x$ direction and
$C_z:=\prod_{i\in C} Z_i$ is a product of flipping operators $Z$
around the closed loop $C$. If $C$ is a homologically trivial loop,
i.e. if it is the boundary of a region on the surface, then $C_z$
can be expressed as a product of $B_p$ operators. Since in $|g.s\ra$
we are summing over all such loops, the action of any $B_p$ on
$|g.s\ra$ leaves it invariant. Since any $A_s$ commutes with all
$B_p$ operators and $A_s|\Omega\ra=|\Omega\ra$, it is immediate that
$A_s|g.s\ra=|g.s\ra$ and hence $|g.s\ra$ is actually a ground state
of (\ref{Hamiltonian}). The other degenerate ground states are
obtained from $|g.s\ra$ by acting on it by product of flipping
operators $Z_i$ around non-trivial homology cycles (Closed loops
which are not boundaries of regions are called homology cycles or
just cycles. On genus one surfaces these are the same as
non-contractible loops, but on higher genus surfaces they are
different.) \\

The excited states are formed by enacting on the ground states by
flipping operators along open strings and hence creating two
particles, \cite{KitaevBasic1, Vidal1} called anyons due to their
exchange properties. While the ground state of (\ref{Hamiltonian})
is rather simple and in fact its entanglement properties and that of
the related topological color codes \cite{Miguel} have been studied
in a number of works \cite{Zanardi1,Kar}, a complete
characterization of the spectrum is difficult, due to the
exponentially large number of open string configurations. In other
words, it is known that any collection of open strings create an
excited state, but determining the degeneracy of such excited states
is not simple, due to the above-mentioned
difficulty. \\

One may expect that on a lattice with a simpler structure, these
problems can be overcome. In this regard, spin ladders  may be of
interest, not only due to their own interest as systems
interpolating between one and two dimensional systems, but also
since they can be used for approximate solution of the more physical
two dimensional systems, when the latter can be
approximated as an array of ladders with negligible couplings between them.\\

In fact the simple structure of ladders facilitates the study of
many interesting phenomena which are otherwise difficult to study in
general lattices. For example one of these phenomena is the dynamics
of defect production, in passing a critical point when a phase
transition occurs \cite{KZ}.  This is usually captured in what is
known as the Kibble-Zurek scaling law. However it has recently been
shown \cite{MiguelPRL}, through a detailed study of a ladder system,
i.e. the Creutz ladder \cite{Cruetz}, that in systems with
topological order, edge states can
dramatically modify this scaling law.  \\

It is the purpose of this paper to make a complete study of the
Kitaev model on spin ladders. The ladder with periodic boundary
condition in one direction has the topology of a cylinder and its
ground state is doubly degenerate. We will determine the energy
spectrum completely and from there we calculate the partition
function and the total entropy. Then we will determine the reduced
density matrices and entropy of various subsystems, where it is
found how the cylinder topology affects these properties. Finally we
discuss briefly how to extend the method to three and higher leg
ladders and eventually to the two dimensional square lattice.

The structure of this paper is as follows. In section (\ref{Model2})
we explain the model and in its various subsections, calculate the
spectrum, the partition function and the thermal averages of
non-local operators. In section (\ref{Ent}) we calculate the density
matrices and entropy of various subsystems which show among other
things that there is no entanglement between any two spins at any
temperature. We end with a discussion on how to extend the method to
larger lattices.

\section{The Kitaev Model on the ladder network}\label{Model2}

\begin{figure}[t]
\centering
\includegraphics[width=12cm,height=2.5cm,angle=0]{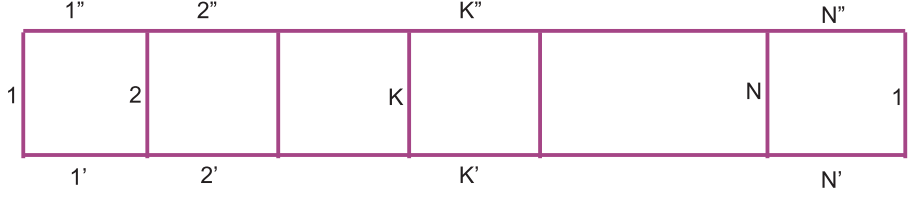}
\caption{The labeling that will be used in the text for the links on
the ladder. }
    \label{BasicLadder}
\end{figure}

Consider a two-leg ladder network with length $N$ with the labeling
of links as shown in figure (\ref{BasicLadder}), where we use the
superscripts $+$ and $-$ to denote the vertices  pertaining to the
lower and upper legs. For this ladder, the vertex and plaquette
operators take the following form,
\begin{equation}\label{AsBp}
    A^+_i:=X_{i'-1}X_iX_{i'},\h A^-_i:=X_{i"-1}X_iX_{i"},\h B_i =
    Z_iZ_{i+1}Z_{i'}Z_{i"},
\end{equation}
Throughout the paper, we use the generic names $A_s$ and $B_p$ for
pointing to general vertex and plaquette operators in a network and
the names indicated in Eq. (\ref{AsBp}) for pointing to specific
operators in the ladder. For a closed surface the operators in
(\ref{Hamiltonian}) are constrained by the relations $\prod_s
A_s=\prod_p B_p=I$, however for the ladder, which is a surface with
two boundaries, the second constraint does not hold and we are left
with
\begin{equation}\label{ProductAs}
\prod_s A_s = I.
\end{equation}
Therefore the operators in (\ref{AsBp}) are $3N-1$ commuting
operators in the $2^{3N}$ dimensional Hilbert space of the ladder.
They all commute with the Hamiltonian and hence the ground state is
two-fold degenerate. The following non-local operators play an
important role.
\begin{equation}\label{WzWx}
    W_z:=\prod_{i'} Z_{i'}\h W_x:=\prod_i X_i,
\end{equation}
It can be clearly seen that they commute with each other and also
with the Hamiltonian,
\begin{equation}\label{Commutation}
    [W_x,W_z]=[H,W_x]=[H,W_z]=0.
\end{equation}
They correspond to a cycle around the ladder. Both of them square to
1 and hence have eigenvalues $\pm 1$. As we will see the operator
$W_z$ generates the two-fold degeneracy of the ground state.

\subsection{The ground and the top states}
It is clear from (\ref{Hamiltonian}) that the ground state is a
state with the property
\begin{equation}\label{GroundStateCondition}
    A_s|\Psi_0\ra=|\Psi_0\ra,\h B_i|\Psi_0\ra = |\Psi_0\ra, \h
    \forall s,\ i.
\end{equation}

To find the explicit form of the ground state, we use the property
$B_i^2=I$ which is equivalent to  $B_i(1+B_i)=1+B_i$ and introduce
the state
\begin{equation}\label{GroundState}
    |\Psi_0\ra:=\frac{1}{\sqrt{2^N}}\prod_i (1+B_i)|\Omega_+\ra,
\end{equation}
where $|\Omega_+\ra:=|+\ra^{\otimes 3N}$, is a product of all spins
in the $|+\ra:=|x,+\ra$ direction. Note that $X|\pm\ra=\pm |\pm\ra\
$ and $ Z|\pm\ra = |\mp\ra$. It is obvious that $|\Psi_0\ra$
satisfies the condition (\ref{GroundStateCondition}) and hence is a
ground state. The other ground state is obtained from $|\Psi_0\ra$
by the action of the operator $W_z$ which corresponds to the only
nontrivial homology cycle of the surface, that is $|\Psi_0\ra$ and
$|\Psi'_0\ra:=W_z|\Psi_0\ra$ form the doubly degenerate ground
states of the model,
\begin{equation}\label{E0}
H(|\Psi_0\ra,|\Psi'_0\ra)=-N(2J+K)(|\Psi_0\ra,|\Psi'_0\ra).
\end{equation}
Note that $W_z$, being a cycle, cannot be expressed as any
combination of product of plaquette operators $B_i$, which are all
homologically trivial. This shows that the two states $|\Psi_0\ra$ and $|\Psi'_0\ra$ are independent. \\

The top state, the state with the highest energy is one which is an
eigenstate of all the vertex and plaquette operators with
eigenvalues $-1$. It can be readily verified that the following
state is such an state:
\begin{equation}\label{TopState}
|\Psi_{top}\ra:=\frac{1}{\sqrt{2^N}}\prod_{i}(1-B_i)|\Omega_-\ra,
\end{equation}
where $|\Omega_-\ra=|-\ra^{\otimes 3N}$ is the product of all spins
in the $|-\ra:=|x,-\ra$ direction in the network. The other
degenerate state is obtained by the action of the operator $W_z$,
i.e. $|\Psi'_{top}\ra=W_z|\Psi_{top}\ra.$ The highest energy will be
$E_{top}=N(2J+K),$
\begin{equation}\label{Etop}
H(|\Psi_{top}\ra,|\Psi'_{top}\ra)=N(2J+K)(|\Psi_{top}\ra,|\Psi'_{top}\ra).
\end{equation}

\subsection{The complete spectrum}\label{Spec}
We can now construct the full spectrum. To this end we note that for
any arbitrary state $|\chi\ra$, the state, $$
    \prod_i
    (1+(-1)^{l_i}B_i)|\chi\ra,\h l_i=\ 0,\ 1,
$$ is an eigenstate of all the plaquette
operators, $B_j$ with eigenvalues $(-1)^{l_j}$.  The reason is the
relation $B_j(1\pm B_j)=\pm (1\pm B_j)$. Hereafter we use the
abbreviation $B_{\bf l}$ for such a string of operators,
\begin{equation}\label{BL}
    B_{{\bf l}}:=  \prod_i
    (1+(-1)^{l_i}B_i).
\end{equation}
These operators satisfy \be\label{BB}B_{{\bf l}}B_{{\bf
l'}}=2^N\delta_{{\bf l,l'}}B_{{\bf l}}.\ee

To construct states which are eigenstates of the vertex operators
and at the same time be independent, let us define the following
operators
\begin{equation}\label{LambdaLRS}
    \Lambda_{{\bf r,s}}:=\prod_{i} Z_i^{r_i}Z_{i'}^{s_i},
\end{equation}
where the sequence of $r_i$ and $s_i$ are 0 or 1 and the labeling
are those shown in figure (\ref{BasicLadder}). Figure (\ref{Lambda2}
) shows the links which contribute to the construction of such
operators. It is important to note that the links of only one leg
are among this set. Now let us define
\begin{equation}\label{ExcitedState}
    |\Psi_{{\bf l,r,s}}\ra :=\Lambda_{{\bf r,s}}|\Psi_{{\bf l}}\ra:=\frac{1}{\sqrt{2^N}}\Lambda_{{\bf
    r,s}}B_{\bf l}|\Omega_+\ra.
\end{equation}
The operators in (\ref{LambdaLRS}) have simple commutation relations
with the vertex and plaquette operators. One can verify the
following relations
\begin{eqnarray}\label{CommutationLambda}
    B_j \Lambda_{{\bf r,s}}&=& \Lambda_{{\bf r,s}}B_j\cr
A^-_j \Lambda_{{\bf r,s}}&=&(-1)^{r_j}\Lambda_{{\bf r,s}}A^-_j\cr
A^+_j \Lambda_{{\bf r,s}}&=&(-1)^{r_j+ s_{j-1}+ s_j}\Lambda_{{\bf
r,s}}A^+_j.
\end{eqnarray}
This leads to
\begin{eqnarray}\label{ABonPsiRLS}
B_j|\Psi_{{\bf l,r,s}}\ra&=&(-1)^{l_{j}}|\Psi_{{\bf l,r,s}}\ra\cr
A^-_j|\Psi_{{\bf l,r,s}}\ra&=&(-1)^{r_j}|\Psi_{{\bf l,r,s}}\ra\cr
A^+_j|\Psi_{{\bf l,r,s}}\ra&=&(-1)^{r_j+ s_{j-1}+ s_j}|\Psi_{{\bf
l,r,s}}\ra.
\end{eqnarray}
The above relations indicate that the states thus constructed are
eigenstates of Hamiltonian,
\begin{equation}\label{HPsi}
H|\Psi_{{\bf l,r,s}}=E_{{\bf l,r,s}}|\Psi_{{\bf l,r,s}}\ra
\end{equation}
where the energy is found from (\ref{Hamiltonian}) and
(\ref{ABonPsiRLS}) to be
\begin{equation}\label{Elrs1}
E_{{\bf l,r,s}}(J,K)=-\sum_j\left[J((-1)^{r_j}+(-1)^{r_j+ s_{j-1}+
s_j})+K(-1)^{l_j}\right].
\end{equation}
In terms of Ising-like variables, $R_i:=(-1)^{r_i}, S_i:=
(-1)^{s_i}, $ and $L_i:= (-1)^{l_i}$ which take values $\pm 1$, the
energy can be rewritten in the form
\begin{equation}\label{Elrs2}
    E_{{\bf
    l,r,s}}=-\sum_j\left[JR_i(S_iS_{i+1}+1)+ KL_i\right].
\end{equation}
In the sequel we will use both the indices $r,s,l$ and $R,S,L$ where
no confusion arises.  The states $|\Psi_{\bf l,r,s}\ra$ have $3N$
binary indices and hence the number of such states is exactly equal
to the dimension of the Hilbert space, hence they will comprise the
full energy spectrum, provided that we can show they are
independent. To investigate this question, let us look at the inner
product of these states.

Using (\ref{CommutationLambda}), and (\ref{ExcitedState}) we find
\begin{eqnarray}\label{ortho}
\la \Psi_{{\bf l,r,s}}|\Psi_{{\bf l', r', s'}}\ra&=&\frac{1}{2^N}\la
\Omega_+|B_{{\bf l}}\Lambda_{{\bf r'},{\bf s'}} \Lambda_{{\bf
r'},{\bf s'}}B_{{\bf l'}}|\Omega_+\ra\cr &=& \delta_{{\bf l,l'}}\la
\Omega_+|B_{{\bf l}}\Lambda_{{\bf r+r'},{\bf s+s'}}|\Omega_+\ra.
\end{eqnarray}
We now note that  the operators $\Lambda_{\bf r+r', s+s'}$ generate
only open or a homologically trivial loop, while the operator
$B_{\bf l}$ only generate homologically trivial loops, and hence the
above matrix element vanishes unless $({\bf r,s})=({\bf r',s'})$.
Moreover by expanding $B_{\bf l}$ it is clearly seen that $\la
\Omega_+|B_{\bf l}|\Omega_+\ra=1$. Hence we find

$$
\la \Psi_{{\bf l,r,s}}|\Psi_{{\bf l', r', s'}}\ra=\delta_{\bf
l,l'}\delta_{\bf r,r'}\delta_{\bf s,s'}$$.

The independence of these states and the equality of their number
with the dimension of Hilbert space, indicates that they comprise
the complete spectrum of the Hamiltonian. It is also instructive to
note the symmetry of the spectrum. If we indicate by ${\bf
\overline{s}}$ the binary complement of the indices ${\bf s}$ (i.e.
$\overline{s}_i=1+s_i$) and similarly for other indices, we find
from (\ref{Elrs1}) that
\begin{eqnarray}\label{Elrs2} E_{{\bf
l,r,\overline{s}}}(J,K)&=&E_{{\bf l,r,s}}(J,K)\cr
E_{{\bf\overline{l},\overline{r},s}}(J,K)&=&E_{{\bf l,r,s}}(-J,-K).
\end{eqnarray}
The first relation expresses the two-fold degeneracy which is the
result of the topology of the surface, that is the action of the
cycle $W_z$ (\ref{WzWx}) on any state, produces another state with
the same energy. The second relation indicates how the spectrum is
affected if we invert the coupling constants $J$ and $K$ around 0;
the spectrum should be inverted around the values $r_j=l_j=0$. One
should also note that only the ground and the top states have
two-fold degeneracy and the degeneracy of the other states is much
larger.

\subsection{The partition function and averages of different
observables}\label{Partition} From the complete spectrum it is
straightforward to calculate the partition function. One writes
\begin{eqnarray}\label{ZBasic}
    Z (\beta,J,K)&=& tr(e^{-\beta H})=\sum_{{\bf L, R, S}}e^{-\beta E_{{\bf
    L,R,S}}}\cr
    &=& \sum_{{\bf R,S}} \prod_j e^{\beta JR_j(S_{j-1}S_j+1)}\sum_{{\bf L}}\prod_j e^{\beta
    KL_j}=:Z_0(\beta,J)Z_1(\beta,K),
\end{eqnarray}
where the last equality defines the partition functions
$Z_0:=Z_0(\beta,J)$ and $Z_1:=Z_1(\beta,K)$. It is obvious that
$Z_1=2^N \cosh^N \beta K$. Using a transfer matrix, we find
\begin{equation}\label{ZBetaJ}
    Z_0=\sum_{{\bf S}} \prod 2 \cosh \beta
    J(S_{j-1}S_j+1) = 2^{2N}(\cosh^{2N} \beta J + \sinh^{2N} \beta J
    ).
\end{equation}
The full partition function is therefore given by
\begin{equation}\label{ZResult}
Z(\beta,J,K)=2^{3N}(\cosh^{2N}\beta J +\sinh^{2N} \beta
J)\cosh^{N}\beta K.
\end{equation}
In the thermodynamic limit, the average energy is obtained from the
partition function to be:
\begin{equation}\label{EAverage}
    \la \frac{E}{N}\ra = -\frac{\partial}{\partial \beta}\ln
    Z(\beta,J,K)= -(2J\tanh \beta J+K\tanh \beta K).
\end{equation}
In the same limit the entropy is found from
$S=(1-\beta\frac{\partial}{\partial \beta})\ln Z$ to be
\begin{equation}\label{Entropy}
    \frac{S}{N} = 3\ln 2+2 \ln \cosh \beta J+\ln \cosh\beta K-2\beta J \tanh 2\beta J-\beta K\tanh \beta
    K.
\end{equation}
The entropy can be written as a sum of two terms, namely
$S=N(S_s(\beta J) + S_p(\beta K))$, where the first one is the
contribution of the vertex terms and the other is the contribution
of plaquette terms, and
\begin{eqnarray}\label{SS}
    S_s(x)&=&2\ln 2 + 2 \ln \cosh x - 2x \tanh 2 x\cr S_p(x)&=&\ln 2 +  \ln \cosh x - x \tanh x.
\end{eqnarray}
Figure (\ref{Entropy}) shows the total entropy as a function of
$\beta J$ and $\beta K$.

\begin{figure}[t]
\centering
\includegraphics[width=10cm,height=10cm,angle=0]{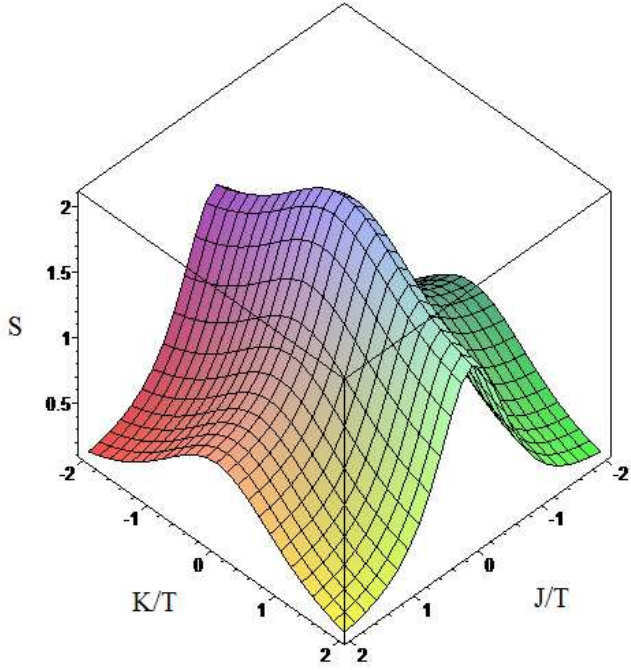}
\caption{Color online. The entropy of the ladder as a function of
the couplings and temperature. }
    \label{Entropy}
\end{figure}

An interesting non-local observable is the string operator $W_x$
defined in (\ref{WzWx}). One finds
\begin{equation}\label{WxAverageBasic}
    \la W_x\ra =  \frac{1}{Z(\beta,J,K)}\sum_{{\bf l,r,s}}\la \Psi_{\bf l}|\Lambda_{\bf r,s}W_x\Lambda_{\bf r,s}|\Psi_{\bf l}\ra e^{-\beta E_{{\bf l,r,s}}}.
\end{equation}
Passing $\Lambda_{\bf r,s}$ through $W_x$, one finds
\begin{equation}\label{WxAverage1}
    \la \Psi_{{\bf l}}|\Lambda_{{\bf r,s}}W_x\Lambda_{{\bf r,s}}|\Psi_{{\bf
    l}}\ra =(-1)^{r_1+r_2+\cdots r_N},
\end{equation}
where we have used the commutation relations of the operators and
also the fact that ($W_xB_j=B_jW_x$\ \ \ $\forall j$) and
consequently $W_x|\Psi_{{\bf l}}\ra=|\Psi_{{\bf l}}\ra.$ Inserting
this into (\ref{WxAverageBasic}) one arrives at
\begin{equation}\label{WxAverage2}
    \la W_x\ra = \frac{1}{Z_0} \sum_{{\bf R, S}}\prod_j
    R_je^{\beta JR_j(S_{j-1}S_j+1)},
\end{equation}
where again an appropriate transfer matrix gives the final result
\begin{equation}\label{WxAverageResult}
    \la W_x\ra = \frac{1}{2^{N-1}}\frac{\sinh ^{N} 2\beta J}{\cosh^{2N} \beta J + \sinh^{2N} \beta J}
\end{equation}
In the thermodynamic limit this gives
\begin{equation}\label{WxThermo}
\la W_x\ra = \left\lbrace\begin{array}{l} 1 \ \ T=0
\\ 0 \ \ T>0\end{array}\right.,
\end{equation}
which shows a phase transition at zero temperature.  It is easy to
see that $\la W_z\ra=0$ at all temperatures.

\section{Reduced density matrices and entropies of different subsystems}\label{Ent}
\begin{figure}[t]
\centering
\includegraphics[width=12cm,height=8.5cm,angle=0]{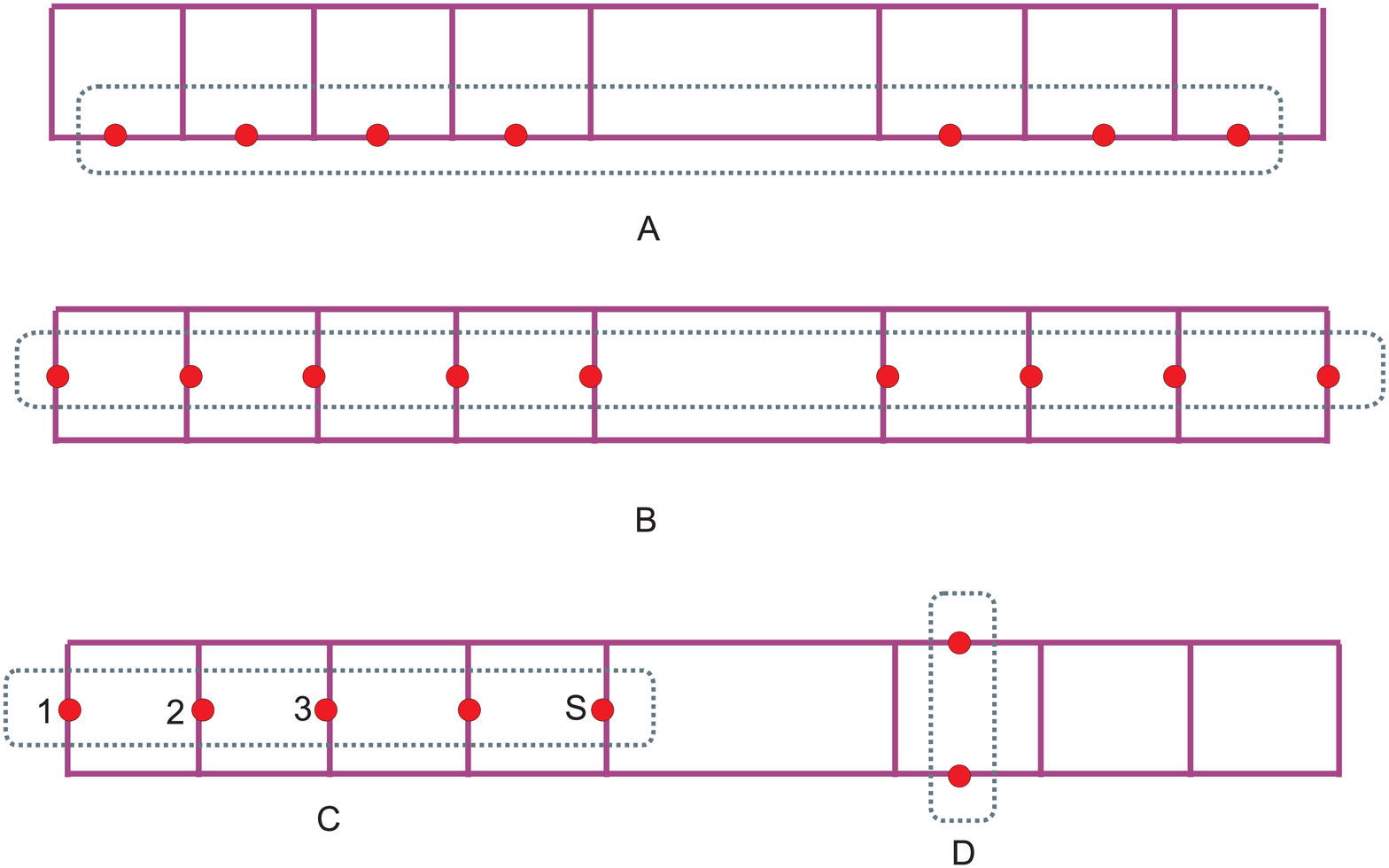}
\caption{Color online. The links with filled circles comprise the
subsystems A, B, C and D. At any temperature, the subsystems A, C
and D are in maximally mixed states. Only the subsystem $B$ which
has a non-trivial topology has a different state, Eq.
(\ref{RhoBTFinal}).}
    \label{Entanglement}
\end{figure}

In this section we derive the reduced density matrices of different
subsystems at finite temperature. These subsystems are denoted by
$A$ (the spins on one of the legs of the ladder), $B$ (the spins of
the totality of all the rungs of the ladder), and $C$ (the spins of
a subset of the rungs), see figure (\ref{Entanglement}). The
significance of the subsystems $B$ and $C$ is that $B$ corresponds
to a topologically nontrivial loop in the surface, while $C$
corresponds to a trivial curve and one expects that this difference
of topology shows itself in the entropy of these subsystems. As we
will show, this is indeed the
case.\\

To prepare ourself for the calculation of the reduced density matrix
at finite temperatures, in each case we first derive the reduced
density matrix of the relevant subsystem  when the whole system is
at the state
 $|\Psi_{{\bf l}}\ra$. The corresponding density matrices will be denoted by $\sigma_A, \sigma_B,$ and $\sigma_C$. This will pave the way for determination of
 the reduced density matrices at arbitrary temperatures which will be denoted by $\rho_A, \rho_B$ and $\rho_C$. We also use the notation $|\Omega_+\ra_A$ to denote the
 restriction of the product state $|\Omega_+\ra$ to the subsystem $A$ with similar notations for $B$ and $C$.  \\
\subsection{Subsystem A: One leg of the ladder}
We have
\begin{eqnarray}\label{RhoABasic}
    \sigma_{_A}:=tr_{\widehat{1",2",\cdots N"}}(|\Psi_{{\bf l}}\ra\la \Psi_{{\bf
    l}}|),
\end{eqnarray}
where for any subset $I$, $\hat{I}$ means that we take the trace
over the complement of $I$.
To calculate the trace, we note that the state $|\Psi_{{\bf l}}\ra$
can be written as follows
\begin{equation}\label{PsiExpand}
    |\Psi_{{\bf l}}\ra = \frac{1}{\sqrt{2^N}}\sum_{m_1,m_2,\cdots
    m_N}(-1)^{\sum_i l_i m_i}B_1^{m_1}\cdots B_N^{m_N}|\Omega_+\ra
\end{equation}
where the indices $m_i$ take the values 0 or 1. We now use the fact
that $B_i=Z_iZ_{i+1}Z_{i'}Z_{i"}$ and take the trace over the
product of the $B_i$ operators to arrive at
\begin{eqnarray}\label{TraceB}
    &&tr_{\widehat{1',2',\cdots N'}}(B_{1}^{m_1}B_2^{m_2}\cdots B_{N}^{m_N}|\Omega_+\ra\la \Omega_+|B_{1}^{k_1}B_2^{k_2}\cdots
    B_{N}^{k_N})\cr &=& Z_{1'}^{m_1}Z_{2'}^{m_2}\cdots
    Z_{{N'}}^{m_N}(\la\Omega_+|M|\Omega_+\ra)_A Z_{1'}^{k_1}Z_{2'}^{k_2}\cdots
    Z_{{N'}}^{k_N},
\end{eqnarray}
where
\begin{equation}\label{M}
    M:=(Z_1Z_2Z_{1"})^{m_1+k_1}(Z_2Z_3Z_{2"})^{m_2+k_2}\cdots.
\end{equation}
From the above equation we find
\begin{equation}\label{M}
   \la\Omega_+|M|\Omega_+\ra_{_A}:= \delta_{{\bf m,\bf k}}=\prod_i
   \delta_{m_i,k_i},
\end{equation}
which when inserted into (\ref{TraceB}) this gives
\begin{eqnarray}\label{RhoAResult}
\sigma_{_A} &=& \frac{1}{2^N} \left(\sum_{m_1,\cdots
m_N}Z_{1'}^{m_1}\cdots Z_{N'}^{m_N}|\Omega_+\ra_A \la \Omega_+|
Z_{1'}^{m_1}\cdots
    Z_{N'}^{m_N}\right) \cr &=& [\frac{1}{2}(\sum_m Z^m|+\ra\la +|Z^m)]^{\otimes
    N}=\frac{1}{2^N}I_A.
\end{eqnarray}
This shows among other things that at zero temperature, each of the
two legs of the ladder are in a maximally mixed state. The
interesting point is that this situation persists at all
temperatures. To see this, we take the upper leg as our subsystem
$A$, since in this case, the analysis will be greatly simplified.
Since the system is symmetric, whatever we obtain will also be valid
for the lower leg. We have \\
\begin{equation}\label{RhoATemperature}
    tr_{\hat{A}}(|\Psi_{{\bf l,r,s}}\ra\la \Psi_{{\bf l,r,s}}|)=tr_{\hat{A}}(\Lambda_{{\bf r,s}}|\Psi_{{\bf l}}\ra\la\Psi_{{\bf l}} |\Lambda_{{\bf r,s}})
    =tr_{\hat{A}}(|\Psi_{{\bf l}}\ra\la \Psi_{_{\bf l}}|) =
    \frac{1}{2^N}I_A,
\end{equation}

The reason for taking $A$ to be the upper leg of the ladder is that
we could pass through  the operator $\Lambda_{\bf r,s}$ cyclically
within the trace and arrive at the simple result that $ \rho_A(T)=
\frac{1}{2^N}I_{A}.$  Obviously any subsystem of $A$ will also be in
a maximally mixed state.

\subsection{Subsystem B: All the rungs of the ladder}

Consider now subsystem $B$, the full set of rungs of the ladder. We
first derive the reduced density matrix $\sigma_B$, when the whole
ladder is in the state $|\Psi_{{\bf l}}\ra$. Using the decomposition
(\ref{PsiExpand}) and the structure of the $B_i $ operators
(\ref{AsBp}), we find
\begin{equation}\label{RhoACalculation2}
    \sigma_{_B}=
    \frac{1}{2^N}\sum_{m_1,\cdots m_N, k_1,\cdots
    k_N}(-1)^{\sum_i l_i(m_i+k_i)}  {_{\hat{B}}}\la\Omega_+|N_{{\bf m,k}}|\Omega_+\ra_{_{\hat{B}}}|\Phi_{{\bf m}}\ra\la \Phi_{{\bf k}}|
\end{equation}
where
\begin{equation}\label{}
    N_{{\bf m,k}} := \prod_i (Z_{i'} Z_{i"})^{m_i+k_i}
\end{equation}
and
\begin{equation}\label{PhiM}
|\Phi_{{\bf m}}\ra:=(Z_1Z_2)^{m_1}\cdots
(Z_{N}Z_1)^{m_N}|\Omega_+\ra_{_B} .
\end{equation}
Using the fact that  ${_{\hat{B}}}\la\Omega_+|N_{{\bf
m,k}}|\Omega_+\ra_{_{\hat{B}}} = \delta_{{\bf m,k}} $, and inserting
the result in (\ref{RhoACalculation2}), noting the two-to-one
correspondence between the indices $m_i$ and the powers of $Z_i$ and
rearranging terms, we obtain
\begin{equation}\label{}
    \sigma_{_B} = \frac{1}{2^{N-1}}\sum_{q_1,\cdots
q_{N-1}}|\tilde{\Phi}_{{\bf q}}\ra\la \tilde{\Phi}_{{\bf q}}|
\end{equation}
where
\begin{equation}\label{PhiM}
|\tilde{\Phi}_{{\bf q}}\ra:=Z_1^{q_1}Z_2^{q_2}\cdots
Z_{N}^{q_1+q_2+\cdots q_{N-1}}|\Omega_+\ra_B
\end{equation}
Note that this is independent of the index set ${\bf l}$ of the
state $|\Psi_{{\bf l}}\ra$. Also in each state $|\tilde{\Phi}_{{\bf
q}}\ra$ the flip operators come in pair, so this state is an even
parity state, i.e. a state where an even number of spins have been
flipped from + to -. The state $\sigma_B$ is thus a uniform mixture
of all even parity states. Call this density matrix $\sigma^{even}$.
Let us now consider finite temperatures, for which we have to
calculate
\begin{equation}\label{PsiBlrs}
    tr_{\widehat{1,2,\cdots N}}(|\Psi_{{\bf l,r,s}}\ra\la \Psi_{{\bf
    l,r,s}}|)= tr_{\widehat{1,2,\cdots N}}(\Lambda_{{\bf r,s}}|\Psi_{{\bf l}}\ra\la \Psi_{{\bf
    l}}|\Lambda_{{\bf r,s}})=\Lambda_{\bf r} \sigma^{(even)} \Lambda_{\bf r},
\end{equation}
where $\Lambda_{\bf r} = \prod_i Z_i^{r_i}$. The reduced density
matrix at finite temperature will now be given by
\begin{equation}\label{RhoBT}
    \rho_{_B}(T) =\frac{1}{Z_0}\sum_{{\bf r,s}}e^{-\beta E_{{\bf
    r,s}}}\Lambda_{\bf r} \sigma^{even} \Lambda_{{\bf r}}.
\end{equation}
From the above definition of $\Lambda_{\bf r}$, and that of
$\sigma^{even}$ and $|\tilde{\Phi}_{\bf q}\ra$, one finds that
\begin{equation}\label{LambdaSigma}
\Lambda_{\bf r}\sigma^{(even)} \Lambda_{\bf r}=
\left\lbrace\begin{array}{l} \sigma^{even} \ ,\  \ |{\bf r}|=0
\\  \sigma^{odd} \ \ , \ \ |{\bf r}|=1,\end{array}\right.
\end{equation}
where $\sigma^{odd}$ is the uniform mixture of odd-parity states and
$|{\bf r}|$ denotes the degree of ${\bf r}$, i.e. $|{\bf
r}|=r_1+r_2+\cdots r_N.$

Inserting this into (\ref{RhoBT}) yields
\begin{equation}\label{RhoBTMiddle}
    \rho_B(T) =\frac{1}{2Z_0}\sum_{{\bf r,s}}e^{-\beta E_{{\bf
    r,s}}} \left((1+(-1)^{|{\bf r}|})\sigma^{even} + (1-(-1)^{|{\bf r}|})\sigma^{odd}\right).
\end{equation}
Using the expression of the parity $|{\bf r}|$ and Eqs.
(\ref{WxAverageBasic}) and (\ref{WxAverage1})
 we find the
following simple expression
\begin{eqnarray}\label{RhoBTFinal}
    \rho_B(T) &=&\frac{1}{2}(1+\la W_x\ra)\sigma^{even}+
    \frac{1}{2}(1-\la W_x\ra)\sigma^{odd}\cr
    &=& \frac{1}{2^N}I_B +
\frac{1}{2}\la W_x\ra (\sigma^{even}-\sigma^{odd}),
\end{eqnarray}
where use has been made of the fact that
$\frac{1}{2}(\sigma^{even}+\sigma^{odd})=\frac{1}{2^N}I_B$. The
entropy of this state, which is a mixture of orthogonal states, can
now be readily calculated. A straightforward calculation gives
\begin{equation}\label{EntropyB}
    S(\rho_B) = N-1 + H(\frac{1+\la W_x\ra}{2}),
\end{equation}
where $H(p)=-p\log_2 p - (1-p)\log_2 (1-p),$ is the Shannon entropy
function.
\begin{figure}[t]
\centering
\includegraphics[width=11cm,height=10cm,angle=0]{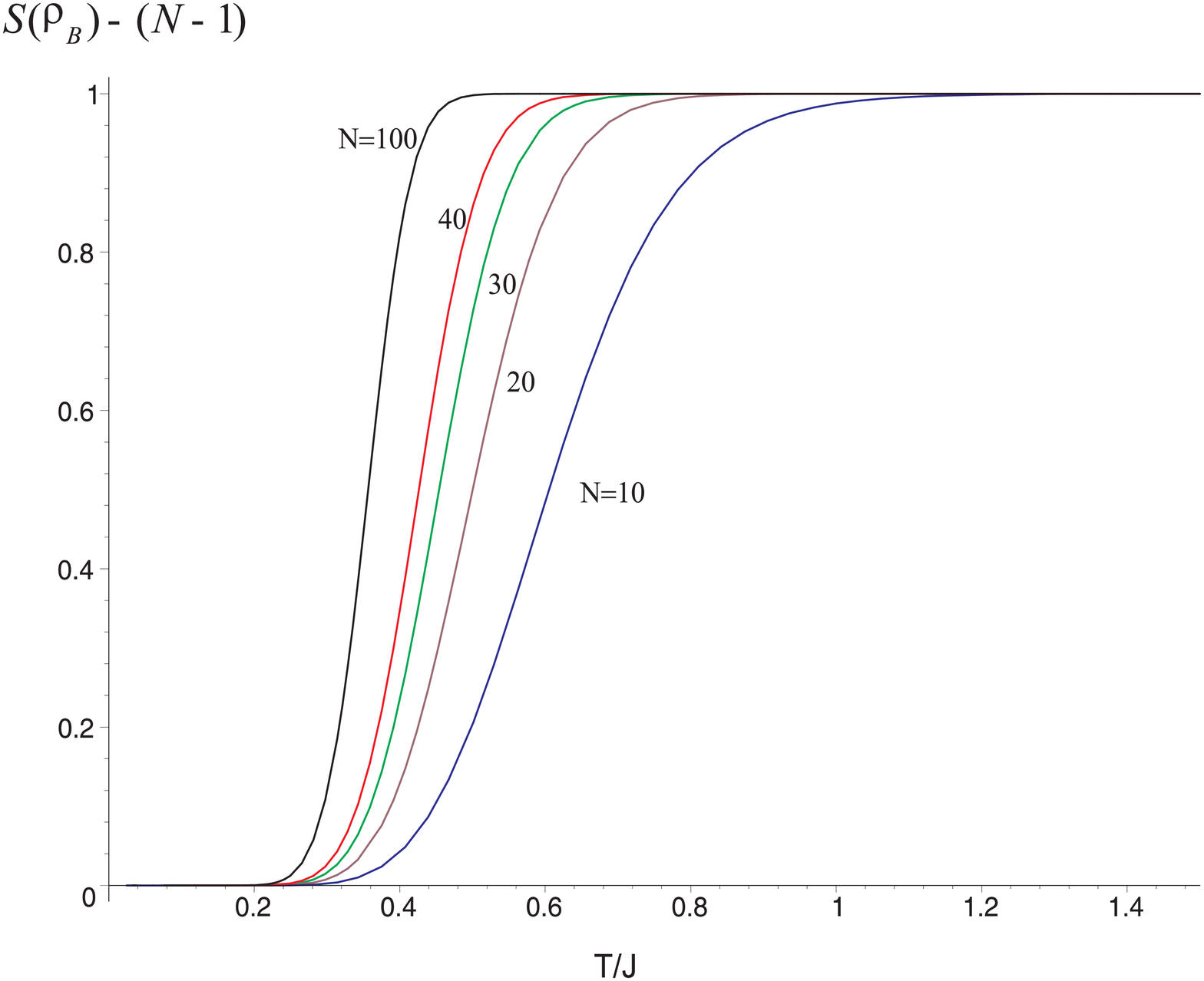}
\caption{Color online. Thermal entanglement of subsystem $B$ with
the rest of the lattice as a function of temperature for different
system sizes. The system sizes are 10, 20, 30, 40 and 100.}
    \label{KitaevCurves}
\end{figure}

Figure (\ref{KitaevCurves}) shows $S_{\rho_B}-(N-1)$ for several
values of system sizes $N$. In the thermodynamic limit, there is a
sharp rise in this quantity only at zero temperature, but for finite
$N$, it is also seen that there is an almost sharp rise at finite
temperatures.

\subsection{Subsystem C: A subset of the rungs}
Once the density matrix of the subsystem B is obtained, we can trace
out any number of the rungs to find the reduced density matrix of
the remaining subset of rungs. Using (\ref{RhoBTFinal}) and noting
that on taking the trace over any subsystem, the contribution of
$\sigma^{even}$ and $\sigma^{odd}$ cancel each other, one arrives at
the simple result that
\begin{equation}\label{C}
    \rho_{_C}=\frac{1}{2^{|C|}} I_C, \h \forall\ \  T.
\end{equation}
where $C$ is any proper subset of the rungs and $|C|$ is the size of
this subset. It is interesting to note that how the topology of the
surface is reflected in the entropy of its subsystems.

\subsection{Thermal entanglement of two spins}
From the reduced density matrices found in previous subsections, we
know that any two spins on a leg of the ladder are in a maximally
mixed state $\rho = \frac{1}{4} I$ and hence there is no thermal
entanglement between such spins. The same is true between any two
spins on the rungs of the ladder. In fact it has been shown
\cite{Zanardi1} that in the ground state, there is no entanglement
between any two qubits. However one can ask if at higher temperature
some degree of entanglement is caused by thermal fluctuations. This
indeed happens in some spin systems, below a certain threshold
temperature. To investigate this, we compute the reduced density
matrices of the two spins, say 1' and 1" in figure
(\ref{Entanglement}) on the two legs, opposite to each other. Call
this subsystem $D$. The first step is to calculate
$tr_{_{\widehat{D}}}|\psi_{\bf l,r,s}\ra\la\psi_{\bf l,r,s} |$ which
is equal to
$$\frac{1}{2^N}tr_{_{\widehat{D}}}\left(\Lambda_{\bf r,s}B_{\bf
l}|\Omega_+\ra\la\Omega_+|\Lambda_{\bf r,s}B_{\bf l}\right)
=\frac{1}{2^N}Z_{1'}^{s_1}tr_{\widehat{_{D}}}\left(B_{\bf
l}|\Omega_+\ra\la\Omega_+|B_{\bf l}\right)Z_{1'}^{s_1} $$. In
calculating the trace, one can use the cyclic property of the trace
and move around all the terms $(1+(-1)^{l_i}B_i)$ except the term
$1+(-1)^{l_1}B_1$ (which acts nontrivially on the space $D$) and use
the property $(1+(-1)^{l_i}B_i)^2=2(1+(-1)^{l_i}B_i)$ which after
some algebra gives
\begin{eqnarray}\label{middleB}
tr_{_{\widehat{D}}}\left(B_{\bf l}|\Omega_+\ra\la\Omega_+|B_{\bf
l}\right)&=& 2^{N-1}tr_{_{\widehat{D}}}\left(\prod_{i\ne
1}(1+(-1)^{l_i}B_i)(1+(-1)^{l_1}B_{1})|\Omega_+\ra\la\Omega_+|(1+(-1)^{l_1}B_{1})\right)
\cr &=&
2^{N-1}tr_{_{\widehat{D}}}\left((1+(-1)^{l_1}B_{1})|\Omega_+\ra\la\Omega_+|(1+(-1)^{l_1}B_{1})\right)
\cr &=&  2^{N-1}\left(|++\ra\la ++|+|--\ra\la --|\right)_{_{1',1"}},
\end{eqnarray}
where in the second line we have used the fact that the closed loops
generated by any of $B_i$'s cannot be compensated by $B_1$ to make a
non-vanishing trace and the third line is the result of explicit
expansion and calculation.  We will then have

\begin{equation}\label{Rho11}
    \rho_{_{D}}=\frac{1}{2Z}\sum_{\bf l,r,s}e^{-\beta E_{\bf
    l,r,s}}Z_{1'}^{s_1}(|++\ra\la ++|+|--\ra\la --|)_{_{1',1"}}Z_{1'}^{s_1}.
\end{equation}
Acting the operators $Z_{1'}^{s_1}$ on both sides and performing the
above simple calculation with the help of the transfer matrix, we
find that $\rho_D=\frac{1}{4}I_D$, which means that there is no
thermal entanglement between these two spins.

\section{The complete spectrum of the three-leg ladder}
\begin{figure}[t]
\centering
\includegraphics[width=12cm,height=3cm,angle=0]{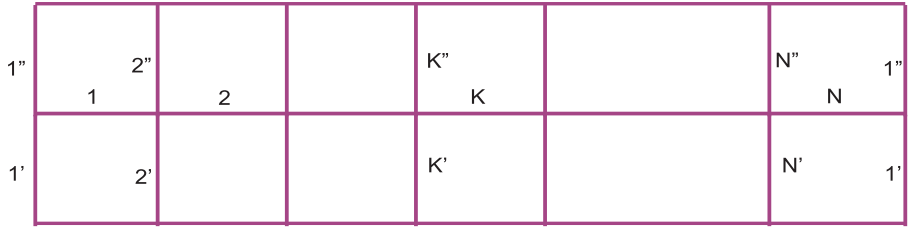}
\caption{The labeling used in the text for the links on the 3-leg
ladder. }
    \label{BasicLadder3}
\end{figure}

\begin{figure}[t]
\centering
\includegraphics[width=10cm,height=5cm,angle=0]{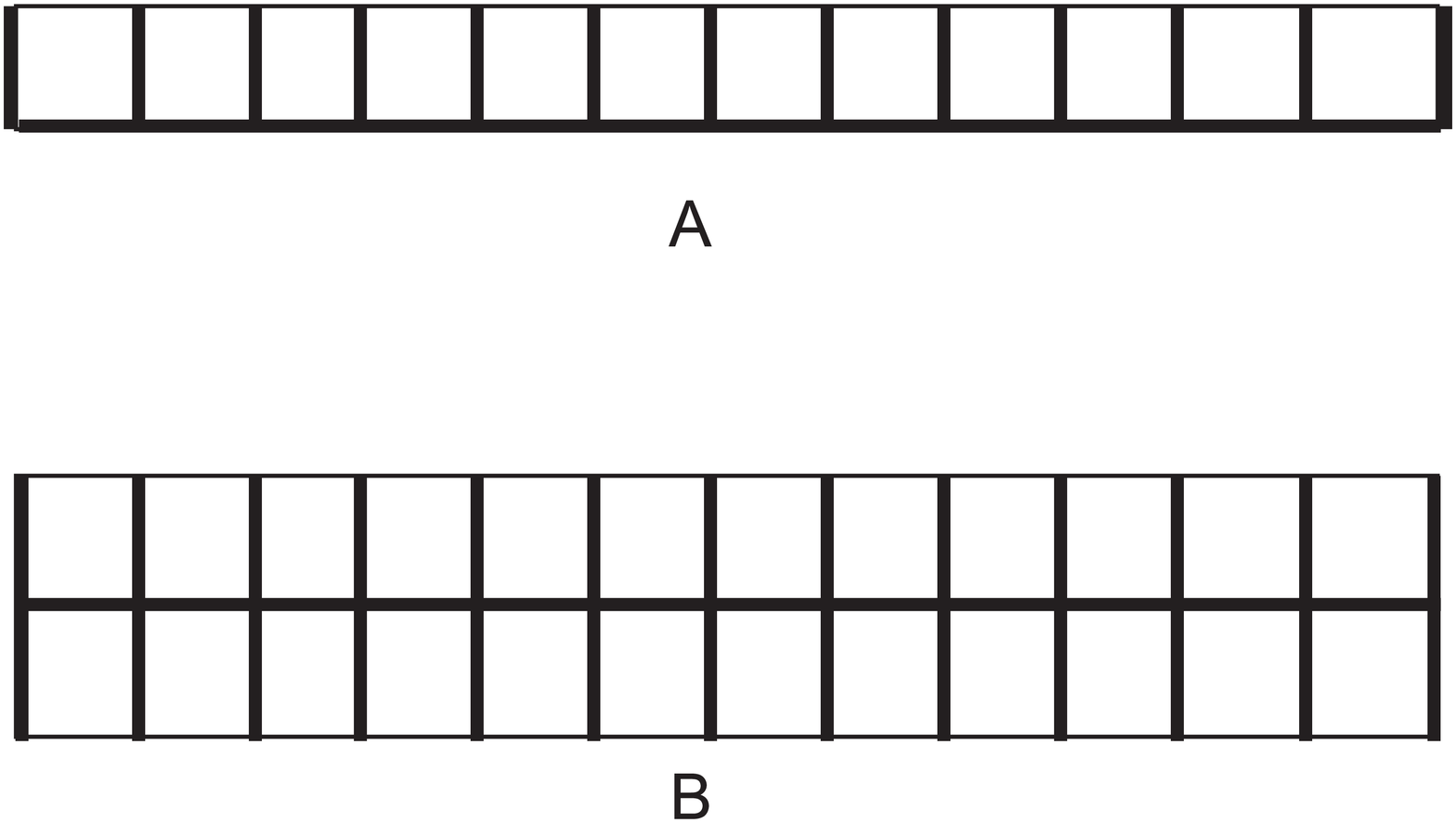}
\caption{The flipping operators are chosen form the set of links
shown in bold, A) for the two-leg ladder, B) for the three-leg
ladder.}
    \label{Lambda2}
\end{figure}

What has been done for the two-leg ladder can be extended to
three-leg ladder, Fig.(\ref{BasicLadder3}),
 without much effort. The first
step is to define the states
\begin{equation}\label{Psi3L}
    |\Psi_{\bf l}\ra = \frac{1}{2^N}\prod_p
    (1+(-1)^{l_p}B_p)|\Omega_+\ra,
\end{equation}
and then the trick is to find a suitable generalization for the
operators $\Lambda_{\bf r,s}$, so that their action on the above
state, produces the correct number of independent eigenstates of the
vertex operators. The suitable generalization is as follows
\begin{equation}\label{LambdaRST}
    \Lambda_{{\bf r,s,t}}:=\prod_{i=1}^N
    Z_i^{r_i}Z_{i'}^{s_i}Z_{i"}^{t_i}
\end{equation}
where the flipping operators correspond to the links shown in Fig.
(\ref{Lambda2}).

One can now easily verify the following commutation relations, where
we use $A_j^+, A_j^0$ and $A_j^-$ for vertex operators on site $j$
for the lower, middle and upper legs of the ladder respectively:
\begin{eqnarray}\label{CommutationLambda3}
A^+_j \Lambda_{{\bf r,s,t}}&=&(-1)^{r_j}\Lambda_{{\bf
r,s,t}}A^+_j\cr A^0_j \Lambda_{{\bf r,s,t}}&=&(-1)^{s_{j-1}+
s_j+r_j+t_j}\Lambda_{{\bf r,s,t}}A^0_j\cr A^-_j \Lambda_{{\bf
r,s,t}}&=&(-1)^{t_j}\Lambda_{{\bf r,s,t}}A^-_j.
\end{eqnarray}

One can proceed along the same way as detailed in section
(\ref{Model2}) for the two-leg ladder and show that the states
$|\Psi_{\bf l,r,s,t}\ra:=\Lambda_{\bf r,s,t}|\Psi_{\bf l}\ra$ are
energy eigenstates with energies given by
\begin{equation}\label{ELRST}
    E=-J\sum_{i}(R_iT_iS_{i-1}S_i+R_i+T_i)-K\sum_i L_i,
\end{equation}
where again we have used Ising-like variables, i.e.
$T_i:=(-1)^{t_i}$ instead of the binary variables. Moreover the
number of these states is $2^{5N}$ which is equal to the dimension
of the Hilbert space and they are orthogonal. (See the reasoning
following Eq.  (\ref{ortho})).

The basic point is that every conceivable combinations of the
flipping operators in $\Lambda_{\bf r,s,t}$, corresponding to Fig.
(\ref{Lambda2}) when acting on $|\Omega_+\ra$, will produce only
open or homologically trivial loops of negative spins which will
certainly be orthogonal to the state generated by $B_{\bf l}$ acting
on $|\Omega_+\ra$. The partition function turns out to be
\begin{equation}\label{Partition3}
    Z(\beta,J,K)= 2^{5N} \cosh^{2N} \beta K\left[\cosh^{3N}\beta J + \sinh^{3N}{\beta J}\right]
\end{equation}
\section{Discussion}

We have determined the complete spectrum of the Kitaev model on a
spin ladder and from there we have determined the reduced density
matrices for its various subsystems  at finite temperature. We have
shown that on two and three-leg ladders, the model is equivalent to
particular types of one dimensional classical Ising models, models
with different spins on the sites and links. \\

To what extent this study can be pursued for the two dimensional
lattice,( i.e. a torus)? On a lattice with $N^2$ sites, and $2N^2$
links, the Hilbert space dimension is $2^{2N^2}$. We can already
construct $2^{N^2}$ (un-normalized) states of the form $|\Psi_{\bf
l}\ra:=\prod_{p}(1+(-1)^{l_p}B_p)|\Omega_+\ra$ which are energy
eigenstates. To find more states, we have to find subsets $I$ of
flipping operators and then construct operators of the form
$\Lambda_{\bf s}:=\prod_{i\in I}Z_i^{s_i}$ and energy eigenstates as
$|\Psi_{\bf s,l}\ra = \Lambda_{\bf s} B_{\bf l}|\Psi_{\bf l}\ra$.
The subset $I$ should have the following important property:  when
acting on $|\Omega_+\ra$, no combination of links in $I$ should be
able to generate a homologically trivial loop of negative spins on
the lattice. Let $I_{max}$ be a maximal set of this type with
$|I_{max}|$ elements. Then the total number of independent energy
eigenstates found in this way is $2^{N^2+|I_{max}|}$. \\

\begin{figure}[t]
\centering
\includegraphics[width=8cm,height=8cm,angle=0]{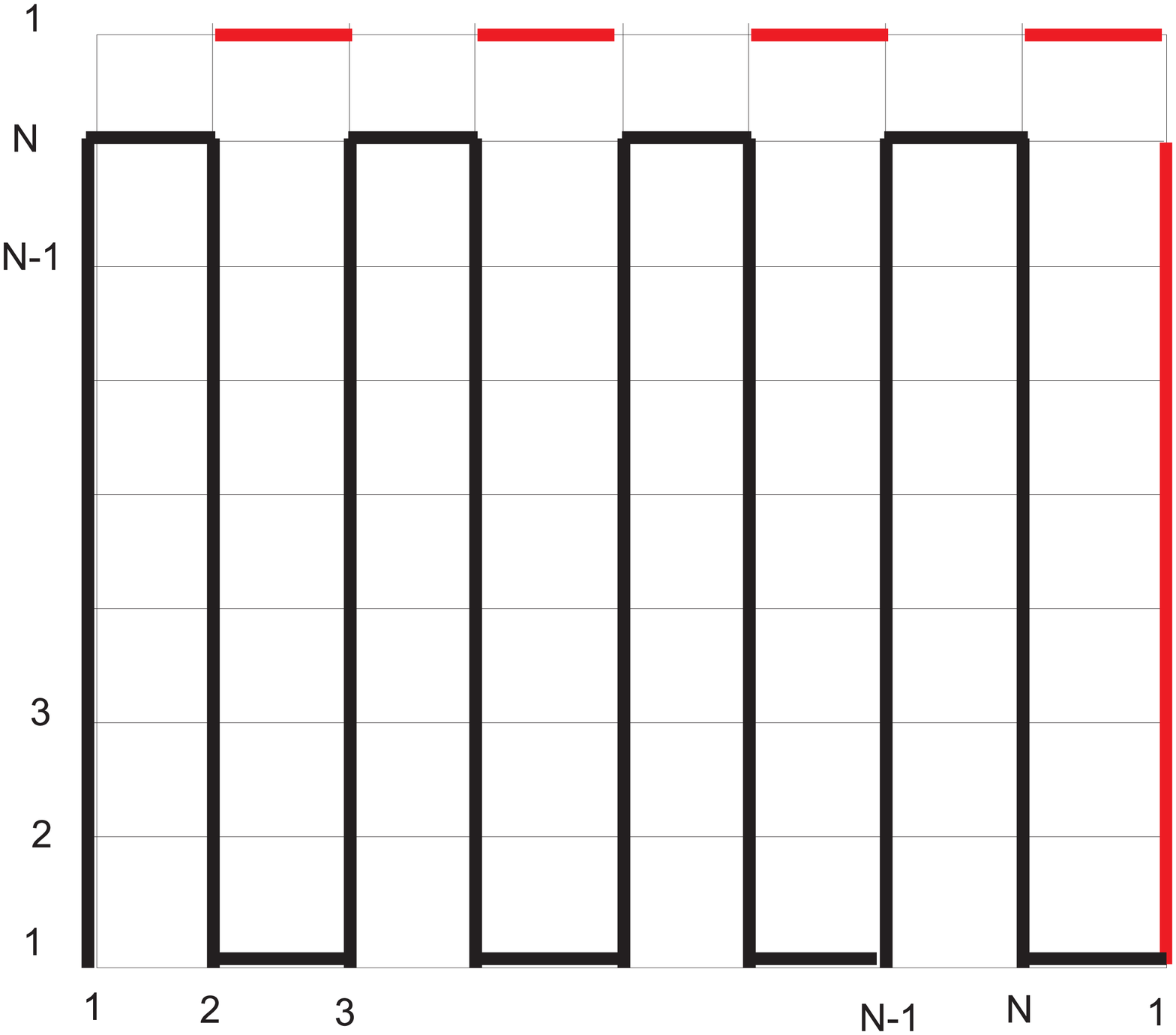}
\caption{Color online. The excitation curve for the two dimensional
lattice, the curve from which the flipping operators are chosen for
creation of the energy eigenstates. The links in red (the ones in
the top and right of the lattice) indicate identical links on the
other sides of the lattice, due to the torus topology.}
    \label{square_5}
\end{figure}

For the square lattice of $N^2$ sites, one such set is shown in Fig.
(\ref{square_5}), where $|I_{max}|=N^2$. In fact $I_{max}$ is
nothing but a one-cycle which goes back and forth around the torus,
but does not wrap around it, and comprises half of the links on the
network. We call the maximal cycle, since it is the cycle which
contains the maximal set of links (the addition of any link to this
cycle will make a trivial loop out of it), or the excitation curve,
since flipping operators chosen from it, and acting on  $|\Psi_{\bf
l}\ra$ create all the excited states. \\

 Since $|I_{max}|=N^2$, the states
constructed as $|\Psi_{\bf m,l}\ra = \Lambda_{\bf m} B_{\bf
l}|\Psi_{\bf l}\ra$ form the whole set of energy eigenstates.
Numbering the links along the bold-face curve shown in figure
(\ref{square_5}), in a consecutive way from $1$ to $N^2$, shows that
the energy of such a state is equal to $E_{\bf
s,l}=-J\sum_{i=1}^{N^2} S_i S_{i+1}-K\sum_i L_i$, where we have used
the Ising type labels $S_i:=(-1)^{s_i}$ and $L_i=(-1)^{l_i}$ instead
of the binary labels $s_i$ and $l_i$. This leads to the partition
function

\begin{equation}\label{Partition3}
    Z_{\rm square\  lattice}(\beta,J,K)= 2^{2N^2} \cosh^{N^2} \beta K\left[\cosh^{N^2}\beta J + \sinh^{N^2}{\beta J}\right]
\end{equation}

{\bf Remark: } We could have taken just such a canonical curve for
the two and three leg ladders, instead of the ones shown in figure
(\ref{Lambda2}), although it may have rendered the calculations of
reduced density matrices in these simple cases unnecessarily
involved. The curves shown in fig (\ref{Lambda2}) can be obtained
from this canonical curve by the moving up and down the horizontal
links appropriately which amounts to application of plaquette
operators. Such plaquette operators affect the eigenstate by only a
phase. \\

Knowing the full spectrum in this way, will enable us to study the
entanglement and many other properties of the Kitaev model in
detail. Furthermore this knowledge may be useful in other more
detailed studies of the Kitaev model, where the dynamics of the
model is required. For example in the study of dynamics of classical
and quantum phase transitions in systems with topological order,
mentioned in the introduction \cite{MiguelPRL}. Another interesting
context is the study of auto-correlation times of the toric code,
which is related to the important problem of how long quantum
information can be protected in topological degrees of freedom in a
background of inevitable thermal fluctuations. This later problem
was first addressed in \cite{ortiz1,ortiz2} in which among other
things, a mapping of the spectrum of the Kitaev model
to two uncoupled Ising chains were found. \\

Another important problem which may be treated in an alternative way
by the characterization of spectrum in the way shown in this paper,
is the problem of sustainment of topological order at finite
temperatures. This problem has been studied in a number of
works,\cite{Castelnovo, aguado1, aguado2} using a different
description of the spectrum.

\section{Acknowledgements} I would like to thank Miguel Martin-Delgado for his very valuable comments and a critical review of the manuscript.

{}
\end{document}